\documentclass[11pt]{article}
\pdfoutput=1
\newcommand{\Comment}[1]{{}}
\usepackage{color}
\usepackage[textwidth = 430 pt, textheight = 630 pt]{geometry}
\definecolor{MyDarkBlue}{rgb}{0.15,0.25,0.45}
\usepackage[linktocpage=true]{hyperref}
\hypersetup{
colorlinks=true,
citecolor=MyDarkBlue,
linkcolor=MyDarkBlue,
urlcolor=MyDarkBlue,
pdfauthor={Constantinos Papageorgakis and Christian S\"amann},
pdftitle={The 3-Lie algebra (2,0) Tensor Multiplets and Equations of Motion on Loop Space},
pdfsubject={hep-th}
breaklinks=true
}

\usepackage[numbers,sort&compress]{natbib}
\usepackage{hypernat,tipa}

\let\fn\footnote
\renewcommand{\footnote}[1]{\linespread{1.1}\fn{#1}\linespread{1.29}}

\usepackage[left]{lineno}

\makeatletter\renewcommand{\section}{\@startsection
{section}{1}{\z@}{-3.5ex plus -1ex minus
    -.2ex}{2.3ex plus .2ex}{\bf }}
\makeatletter\renewcommand{\subsection}{\@startsection{subsection}{2}{\z@}{-3.25ex
plus -1ex minus
   -.2ex}{1.5ex plus .2ex}{\it }}
\makeatletter\renewcommand{\subsubsection}{\@startsection{subsubsection}{3}{-2.45ex}{-3.25ex
plus -1ex minus -.2ex}{1.5ex plus .2ex}{\it }}

\renewcommand{\@seccntformat}[1]{\@nameuse{the#1}.~~}

\makeatletter \@addtoreset{equation}{section}

\usepackage{epsfig,rotating}
\usepackage{amsmath,amssymb}
\usepackage{amsfonts}
\usepackage{mathrsfs}
\usepackage{bbm}
\usepackage{bm}

\renewcommand{\slash}[1]{#1\hspace{-0.27cm}/\,}

\def\periodb#1{\setbox0=\hbox{$#1$}#1\hskip-\wd0\hbox to\wd0{-}}
\newcommand{\nablas}{\slash{\nabla}}
\newcommand{\nablabs}{\slash{\bar{\nabla}}}

\newcommand{\xd}{\dot{x}}
\newcommand{\cA}{\mathring{A}}
\newcommand{\cF}{{\mathring{F}}}
\newcommand{\cX}{{\mathring{X}}}
\newcommand{\cPhi}{{\mathring{\Phi}}}
\newcommand{\cPsi}{{\mathring{\Psi}}}

\newcommand{\lbr}{(\hspace{-0.1cm}(}
\newcommand{\rbr}{)\hspace{-0.1cm})}

\newcommand{\unit}{\mathbbm{1}}   			
\renewcommand{\span}{\mathrm{span}}   			

\newcommand{\CA}{\mathcal{A}}    			

\newcommand{\CB}{\mathcal{B}}

\newcommand{\CI}{\mathcal{I}}

\newcommand{\CL}{\mathcal{L}}

\newcommand{\CN}{\mathcal{N}}

\newcommand{\CP}{\mathcal{P}}

\newcommand{\CT}{\mathcal{T}}

\newcommand{\CU}{\mathcal{U}}

\newcommand{\frg}{\mathfrak{g}}				

\newcommand{\FR}{\mathbbm{R}}     			
\newcommand{\FC}{\mathbbm{C}}     			

\newcommand{\dd}{\mathrm{d}}     			
\newcommand{\dpar}{\partial}     			
\newcommand{\di}{\mathrm{i}}     			
\newcommand{\eps}{{\varepsilon}}			
\newcommand{\epsb}{{\bar{\varepsilon}}}			

\newcommand{\psib}{{\bar{\psi}}}

\newcommand{\eand}{{~~~\mbox{and}~~~}}     		

\newcommand{\der}[1]{\frac{\dpar}{\dpar #1}}   		
\newcommand{\dder}[1]{\frac{\dd}{\dd #1}}   		

\newcommand{\au}{\mathfrak{u}}
\newcommand{\asu}{\mathfrak{su}}
\newcommand{\aso}{\mathfrak{so}}

\newcommand{\sU}{\mathsf{U}}     			

\newcommand{\sSU}{\mathsf{SU}}

\newcommand{\sSO}{\mathsf{SO}}

\newcommand{\sSpin}{\mathsf{Spin}}

\newcommand{\ie}{{\it i.e.}}
\newcommand{\eg}{{\it e.g.}}

\newcommand{\acton}{\vartriangleright}     			
\newcommand{\remark}[1]{}     				

\newcommand{\dotsp}{\;\cdot\;}
     
\def\tyng(#1){\hbox{\tiny$\yng(#1)$}}			
\def\tyoung(#1){\hbox{\tiny$\young(#1)$}}			

\begin{document}

\renewcommand{\thefootnote}{\fnsymbol{footnote}}

\rightline{KCL-MTH--11--06}
\rightline{HWM--11--06}
\rightline{EMPG--11--07}

   \vspace{1.8truecm}

\vspace{15pt}

\centerline{\LARGE \bf {\sc The 3-Lie algebra (2,0) Tensor Multiplet}}\vspace{.5cm} \centerline{\LARGE \bf {\sc and Equations of Motion on Loop Space}} \vspace{2truecm} \thispagestyle{empty} \centerline{
    {\large {\bf{\sc Constantinos Papageorgakis${}^{\,a,}$}}}\footnote{E-mail address:
                                 \href{mailto:costis.papageorgakis@kcl.ac.uk}{\tt costis.papageorgakis@kcl.ac.uk}} {and} {\large {\bf{\sc Christian S\"amann${}^{\,b,}$}}}\footnote{E-mail address:
                                 \href{mailto:C.Saemann@hw.ac.uk}{\tt c.saemann@hw.ac.uk} }
                                                           }

\vspace{1cm}
\centerline{${}^a${\it Department of Mathematics, King's College London}}
\centerline{{\it The Strand, London WC2R 2LS, UK}}
\vspace{.8cm}
\centerline{${}^b${\it Department of Mathematics,
Heriot-Watt University}}
\centerline{{\it Colin Maclaurin Building, Riccarton, Edinburgh EH14 4AS, UK}}
\centerline{{\it and}}
\centerline{{\it Maxwell Institute for Mathematical Sciences, Edinburgh,
UK}}

\vspace{2.0truecm}

\thispagestyle{empty}

\centerline{\sc Abstract}

\vspace{0.4truecm}
\begin{center}
\begin{minipage}[c]{380pt}{
    \noindent We show that a recently found set of supersymmetric equations of motion for a 3-Lie algebra-valued (2,0) tensor multiplet finds a natural interpretation as supersymmetric gauge field equations on loop space. We find that BPS solutions to these equations yield a previously proposed nonabelian extension of the selfdual string. We describe an ADHMN-like construction that allows for the explicit construction of such BPS solutions.}
\end{minipage}
\end{center}

\vspace{.4truecm}

\noindent

\vspace{.5cm}

\setcounter{page}{0}
\parskip 5pt

\newpage

\renewcommand{\thefootnote}{\arabic{footnote}}
\setcounter{footnote}{0}

\section{Introduction and summary of results}

There has been recent success in formulating Lagrangian descriptions for multiple M2-branes in terms of the 3-Lie algebra theories of Bagger-Lambert and Gustavsson (BLG) \cite{Bagger:2006sk,Gustavsson:2007vu,Bagger:2007jr,Bagger:2007vi,Bagger:2008se}, as well as the closely related Chern-Simons-matter theories of Aharony-Bergman-Jafferis-Maldacena (ABJM) \cite{Aharony:2008ug}. In this context, it is natural to ask whether any of these results extend to the case of multiple M5-branes. An attempt to find a 3-Lie algebra theory in six dimensions with (2,0) supersymmetry was made in \cite{Lambert:2010wm}. There, it was shown that the supersymmetry algebra closes on-shell after introducing an auxiliary gauge field and a covariantly constant vector $C^\mu$, $\mu = 0,\ldots,5$, in addition to the expected field content of the nonabelian tensor multiplet. However, the resulting theory reduces to five-dimensional super Yang-Mills (SYM) theory when expanded around any nonzero vacuum solution for $C^\mu$ and no inherently six-dimensional dynamics were found.\footnote{The theory expanded around the vacuum $\langle C^\mu \rangle = 0$ led to a collection of free tensor multiplets.} The interpretation of the full theory remains obscure. For subsequent applications see \cite{Kawamoto:2011ab,Honma:2011br}.

On a different track, there have also been alternative past attempts at finding an M5-brane theory by employing loop spaces. For the abelian case it has been argued that the 2-form potential $B_{\mu\nu}$ can be interpreted as a connection on the bundle of all loops in spacetime \cite{Freund:1981qw}. This is due to the existence of a so-called transgression map \cite{0817647309}: On loop space, there is a natural vector corresponding to the tangent vector to a loop. This vector can be used to lower the degree of a differential form by contraction. In particular, this operation allows us to translate the curvature 3-form of a gerbe to a 2-form, which can be interpreted as the curvature of a gauge bundle. This map has been successfully used in a lift of the ADHMN construction to a construction of selfdual string solitons in M-theory  \cite{Saemann:2010cp}. It is therefore natural to ask for a loop space description of a six-dimensional theory with (2,0) supersymmetry. Proposals along these lines can be found in \cite{Gustavsson:2005fp,Gustavsson:2005aq,Gustavsson:2008dy}, see also \cite{Kawamoto:2000zt,Bergshoeff:2000jn}, where noncommutative loop spaces were derived.

In this paper, we combine these two sets of ideas by interpreting the 3-algebra (2,0) theory of \cite{Lambert:2010wm} as a supersymmetric theory on loop space. This approach is motivated by the observation that the equation of motion for the gauge field strength found in \cite{Lambert:2010wm} is indeed very similar to a transgression, as we will show. A first attempt at connecting \cite{Lambert:2010wm} with a theory in loop space was made in \cite{Huang:2010db} but remained unsuccessful in obtaining closure of the supersymmetry algebra.

Our transition to loop space is implemented by an extended transgression map between a 3-Lie algebra theory with a selfdual 3-form field-strength $H$ on $\FR^{1,5}$ and a gauge theory with a 2-form field-strength $\cF$ on the loop space $\mathcal L \mathcal \FR^{1,5}$ of $\FR^{1,5}$. In the resulting equations of motion, which are summarized in \eqref{eq:eomsummary}, the selfduality of the 3-form field disappears. However, other constraints found in the 3-Lie algebra picture still have to be imposed on loop space. After fixing loops to wrap a compactified `M-theory direction', the equations of motion on loop space are easily seen to be equivalent to SYM theory in five dimensions, as one would expect from \cite{Lambert:2010wm}.

We then move on to studying the BPS sector of our equations: In \cite{Saemann:2010cp}, solutions to the Basu-Harvey equation \cite{Basu:2004ed} were used to construct selfdual string solitons of an abelian M5-brane theory.\footnote{For another loop space approach to the selfdual string soliton see \cite{Gustavsson:2006ie,Gustavsson:2008dy}.} This procedure followed the steps of the ADHMN construction \cite{Nahm:1979yw,Nahm:1981nb,Nahm:1982jb}: Solutions to the Basu-Harvey equation yield a twisted Dirac operator, whose zero modes lead to fields which solve the transgressed form of the selfdual string equation on loop space. This loop space version of the selfdual string equation allows for a natural nonabelian generalization suggested in \cite{Saemann:2010cp}. Interestingly, we find that this generalization is indeed the appropriate BPS equation of our loop space version of the supersymmetric 3-Lie algebra (2,0) tensor multiplet equations. Moreover, we manage to extend the ADHMN-like construction of \cite{Saemann:2010cp} to this nonabelian case and perform a simple example of such a construction explicitly. 

This paper is organized as follows: In Section \ref{equations}, we present some facts about 3-Lie algebras and we review the supersymmetric equations of motion for the 3-Lie algebra (2,0) tensor multiplet found in \cite{Lambert:2010wm}. In Section \ref{gerbes}, we discuss the reinterpretation of these equations in terms of gauge field equations on loop space. In Section \ref{nonabeliansds}, we study the BPS sector of these equations and give the extended ADHMN-construction. We conclude in Section \ref{sec:conclusions}.

\section{Nonabelian tensor multiplet equations}\label{equations}

In \cite{Lambert:2010wm}, the method originally used to derive the BLG model \cite{Bagger:2007jr} was applied to find suitable nonabelian equations for the six-dimensional tensor multiplet. That is, the closure of the algebra of certain postulated supersymmetry transformations, which was made possible by introducing an additional gauge potential and a vector field, yielded the desired equations. Below, we briefly review these tensor multiplet equations.

\subsection{3-Lie algebras}

A {\em 3-Lie algebra} \cite{Filippov:1985aa} is a vector space $\CA$ endowed with a totally antisymmetric, trilinear map $[\dotsp,\dotsp,\dotsp]:\CA^{\wedge 3}\rightarrow \CA$ which satisfies the {\em fundamental identity}
\begin{equation}\label{eq:fundamentalIdentity}
 [a,b,[x,y,z]]=[[a,b,x],y,z]+[x,[a,b,y],z]+[x,y,[a,b,z]]~,~~~a,b,x,y,z\in\CA~.
\end{equation}
We demand that $\CA$ is endowed with a nondegenerate invariant symmetric bilinear form, \ie\ a nondegenerate map $(\dotsp,\dotsp):\CA\odot\CA\rightarrow \FC$ satisfying the {\em compatibility condition}
\begin{equation}
 ([a,b,x],y)+(x,[a,b,y])=0
\end{equation}
for all $a,b,x,y\in\CA$. Therefore, $\CA$ is in fact a {\em metric 3-Lie algebra}. Any such 3-Lie algebra $\CA$ comes with an {\em associated Lie algebra} $\frg_\CA$ of inner derivations, which is spanned as a vector space by the maps $D(a,b):\CA\rightarrow \CA$, $a,b\in\CA$, defined via 
\begin{equation}
 D(a,b)\acton x:=[a,b,x]~,~~~x\in \CA~.
\end{equation}
Note that the Lie bracket $[X,Y]$, $X,Y\in\frg_\CA$, closes due to the fundamental identity \eqref{eq:fundamentalIdentity}. On $\frg_\CA$, there is the invariant symmetric bilinear form
\begin{equation}\label{eq:invSymForm}
 \lbr D(a,b),D(c,d)\rbr=([a,b,c],d)~,~~~a,b,c,d\in\CA~,
\end{equation}
which is induced from the invariant form on the 3-Lie algebra. The only nontrivial 3-Lie algebra with positive definite metric is $A_4$,\footnote{One can also consider direct sums of $A_4$.} which corresponds to the vector space $\FR^4$ with Euclidean standard basis $(e_1,\ldots,e_4)$ together with the 3-bracket and metric
\begin{equation}
 [e_i,e_j,e_k]=\eps_{ijkl}e_l\eand (e_i,e_j)=\delta_{ij}~.
\end{equation}
Here, the associated Lie algebra is  $\frg_{A_4}=\aso(4)\cong\asu(2)\oplus\asu(2)$, and the invariant form \eqref{eq:invSymForm} on $\frg_{A_4}$ is of split signature. For further details on 3-Lie algebras, see \eg\ \cite{deAzcarraga:2010mr} and references therein.

\subsection{The 3-Lie algebra tensor multiplet equations}

The field content of a tensor multiplet is given by five scalar fields $X^I$, a selfdual 3-form field strength $H_{\mu\nu\kappa}$,
\begin{equation}
 H_{\mu\nu\kappa}=\tfrac{1}{3!}\eps_{\mu\nu\kappa\rho\sigma\tau}H^{\rho\sigma\tau}~,
\end{equation}
and the corresponding superpartners, which are here combined into a Majorana spinor of $\sSO(1,10)$ satisfying $\Gamma_{012345}\Psi=-\Psi$. In the following, it is assumed that all these fields take values in a 3-Lie algebra $\CA$. To allow the supersymmetry algebra to close, it was found that a covariantly constant vector $C^\mu$, with values in $\CA$, and an additional gauge potential $A_\mu$, living in the associated Lie algebra $\frg_\CA$ of $\CA$, had to be introduced. The latter yields covariant derivatives acting on the matter fields according to
\begin{equation}
 \nabla_\mu X^I=\dpar_\mu X^I+ A_\mu\acton X^I=\dpar_\mu X^I+ A_\mu^{ab}D(\lambda_a,\lambda_b)\acton X^I=\dpar_\mu X^I+ A_\mu^{ab}[\lambda_a,\lambda_b,X^I]~,
\end{equation}
where $\lambda_a$ denotes the generators of the 3-Lie algebra $\CA$. Altogether, the equations of motion for a 3-Lie algebra-valued (2,0) tensor multiplet found in \cite{Lambert:2010wm} read as
\begin{equation}\label{eq:eomTensor}
\begin{aligned}
 \nabla^2 X^I-\tfrac{\di}{2}[\bar{\Psi},\Gamma_\nu\Gamma^I\Psi,C^\nu]+[X^J,C^\nu,[X^J,C_\nu,X^I]]&=0~,\\
 \Gamma^\mu\nabla_\mu\Psi-[X^I,C^\nu,\Gamma_\nu\Gamma^I\Psi]&=0~,\\
 \nabla_{[\mu}H_{\nu\kappa\lambda]}+\tfrac{1}{4}\eps_{\mu\nu\kappa\lambda\sigma\tau}[X^I,\nabla^\tau X^I,C^\sigma]+\tfrac{\di}{8}\eps_{\mu\nu\kappa\lambda\sigma\tau}[\bar{\Psi},\Gamma^\tau\Psi,C^\sigma]&=0~,\\
F_{\mu\nu}-D(C^\lambda,H_{\mu\nu\lambda})&=0~,\\
\nabla_\mu C^\nu&=0~,\\
D(C^\mu,C^\nu)&=0~,\\
D(C^\rho,\nabla_\rho X^I)=D(C^\rho,\nabla_\rho\Psi)=D(C^\rho,\nabla_\rho H_{\mu\nu\lambda})&=0~.
\end{aligned}
\end{equation}
In our notation, the 3-bracket acts exclusively onto the 3-Lie algebra structure. That is, \eg\,
\begin{equation}
 [\bar{\Psi},\Gamma^\tau\Psi,C^\sigma]=\bar{\Psi}^a\Gamma^\tau\Psi^bC^{\sigma c}[\lambda_a,\lambda_b,\lambda_c]~.
\end{equation}
The supersymmetry transformations under which these equations are invariant read as \cite{Lambert:2010wm}
\begin{equation}
 \begin{aligned}
  \delta X^I&=\di \epsb \Gamma^I\Psi~,\\
  \delta \Psi&=\Gamma^\mu\Gamma^I\nabla_\mu X^I\eps+\tfrac{1}{2\times 3!}\Gamma_{\mu\nu\lambda}H^{\mu\nu\lambda}\eps-\tfrac{1}{2}\Gamma^{IJ}\Gamma_\lambda[X^I,X^J,C^\lambda]\eps~,\\
  \delta H_{\mu\nu\lambda}&=3\di \epsb\Gamma_{[\mu\nu}\nabla_{\lambda]}\Psi+\di\epsb\Gamma^I\Gamma_{\mu\nu\lambda\kappa}[X^I,\Psi,C^\kappa]~,\\
  \delta A_\mu&=\di\epsb\Gamma_{\mu\lambda} D(C^\lambda,\Psi)~,\\
  \delta C^\mu&=0~,
 \end{aligned}
\end{equation}
where the Majorana spinor $\eps$ satisfies the additional condition $\Gamma_{012345}\eps=\eps$. 

\section{3-Lie algebra tensor multiplet equations on loop space}\label{gerbes}

We now come to our reinterpretation of the tensor multiplet equations \eqref{eq:eomTensor} as gauge theory equations on loop space. For this, we first recall the relation between gerbes and gauge bundles on loop space, before extending this relation to the 3-Lie algebra (2,0) tensor multiplet.

\subsection{Abelian gerbes}

In the following, we adopt the conventions and notations of \cite{Saemann:2010cp}. We are exclusively working with {\em abelian local} or {\em Hitchin-Chatterjee gerbes}. 

Consider a principal $\sU(1)$-bundle $P$ over a manifold $M$, which comes with an open cover $\CU=(U_i)$. The structure of the principal fiber bundle can be encoded in a $\au(1)$-valued connection 1-form. Its curvature is globally defined, while the corresponding gauge potential is defined on the patches $U_i$. The transition function $f_{(ij)}$ from patch $U_j$ to patch $U_i$ is reconstructed by comparing gauge potentials on different patches. Repeated application of the Poincar{\'e} lemma yields the formulas
\begin{equation}
 F=\dd A_{(i)}~~\mbox{on}~~U_i\eand A_{(i)}-A_{(j)}=\dd \log f_{(ij)}~~\mbox{on}~~U_i\cap U_j~.
\end{equation}

Abelian gerbes are obtained by shifting this picture by one degree in the cohomology. That is, we have a globally defined 3-form $H$, 2-forms $B_{(i)}$ on the patches $U_i$, 1-forms $A_{(ij)}$ on intersections $U_i\cap U_j$ and functions $h_{(ijk)}$ on triple intersections $U_i\cap U_j\cap U_k$, which are related via
\begin{equation}
\begin{aligned}
 H=\dd B_{(i)}~\mbox{on}~U_i~,~~~B_{(i)}-B_{(j)}=\dd A_{(ij)}~\mbox{on}~U_i\cap U_j~,\\
A_{(ij)}-A_{(ik)}+A_{(jk)}=\dd h_{(ijk)}~\mbox{on}~U_i\cap U_j\cap U_k~.~~~~
\end{aligned}
\end{equation}
All the fields take values in $\au(1)$ and together, they form the {\em connective structure} of a gerbe.\footnote{We can also define a $\sU(1)$-valued function on triple intersections by $f_{(ijk)}:=\exp(\di h_{(ijk)})$.} As we obtained a gauge potential on intersections, we can also think of abelian local gerbes as principal $\sU(1)$-bundles defined on the intersections of patches endowed with further compatibility conditions.

\subsection{Transgression of the connective structure}

Consider a manifold $M$, its {\em loop space} $\CL M:={\rm Map}(S^1,M)$ and the {\em correspondence space} $\CL M\times S^1$. There is an obvious projection from the correspondence space to the loop space. Furthermore, a point in the correspondence space can be mapped to a point on $M$ by the evaluation map $ev$. This map projects a loop $x(\tau)$, $\tau\in[0,2\pi)$, and an angle $\tau_0\in S^1$ to the point $x(\tau_0)$ on the loop corresponding to the angle $\tau_0$. Altogether, we arrive at the following double fibration
\begin{equation}\label{dblfibrationfourself}
\begin{aligned}
\begin{picture}(50,40)
\put(0.0,0.0){\makebox(0,0)[c]{$M$}}
\put(64.0,0.0){\makebox(0,0)[c]{$\CL M$}}
\put(32.0,33.0){\makebox(0,0)[c]{$\CL M\times S^1$}}
\put(7.0,18.0){\makebox(0,0)[c]{$ev$}}
\put(62.0,18.0){\makebox(0,0)[c]{$pr$}}
\put(25.0,25.0){\vector(-1,-1){18}}
\put(37.0,25.0){\vector(1,-1){18}}
\end{picture}
\end{aligned}
\end{equation}
We can now construct a {\em transgression map} $\CT:\Omega^{k+1}(M)\rightarrow \Omega^k(\CL M)$ such that $\CT=(pr)_!\circ ev^*$, cf.\ \cite{0817647309}. That is, given a differential form $\omega\in\Omega^k(M)$, $k>0$, the transgression map pulls it back along $ev$ and pushes it forward\footnote{The existence of this map is nontrivial.} along $pr$. Explicitly, the transgression map is defined by
\begin{equation}\label{def:transgression}
 (\CT\omega)_x(v_1(x),\ldots,v_k(x)):=\oint_{S^1}\dd \tau\,\omega(v_1(\tau),\ldots,v_k(\tau),\dot{x}(\tau))~,
\end{equation}
where $x\in\CL M$, and the $v_i(x)\in T \CL M$ are vector fields on $\CL M$. We see that the existence of a preferred vector on loop space allows us to reduce a form degree by contraction. The loop spaces we are interested in consist of loops $x(\tau)$ with a fixed choice of parametrization such that $|\xd(\tau)|=R>0$, with $2\pi R$ being the length of the loop. This excludes loops with timelike or lightlike tangent vectors. 

For the connective structure of a gerbe, this means that the curvature 3-form $H$ is mapped to a 2-form gauge field strength $F$. One has therefore the freedom to either work on a finite-dimensional base space endowed with an abelian gerbe or with an ordinary principal $\sU(1)$-bundle over an infinite-dimensional manifold given by the loop space of the original manifold. Note that a transgression can be shown to be invertible up to gauge transformations and therefore corresponds to a mere rewriting.

Note also that in Cartesian coordinates on Minkowski space $\FR^{1,5}$, the transgression map \eqref{def:transgression} reads as
\begin{equation}\label{eq:full}
 (\CT\omega)_x(v_1(x),\ldots,v_k(x))=\oint_{S^1}\dd \tau\,\omega_{\mu_1\ldots\mu_k\mu_{k+1}}v_1^{\mu_1}(\tau)\ldots v_k^{\mu_k}(\tau)\dot{x}^{\mu_{k+1}}(\tau)~.
\end{equation} 
It is in one-to-one correspondence with the expression $\omega_{\mu_1\ldots\mu_k\mu_{k+1}}\dot{x}^{\mu_{k+1}}(\tau)$. We can therefore drop the integral over the angle $\tau$ in equations concerning transgressed fields as done in \cite{Saemann:2010cp}, which will simplify our notation.

\subsection{The conditions on the auxiliary field $C^\mu$}

The equation for the 3-form and gauge field strengths are very reminiscent of an extended 3-Lie algebra transgression to the loop space $\CL\FR^{1,5}$:
\begin{equation}\label{eq:eomHF}
 F_{\mu\nu}=D(C^\lambda,H_{\mu\nu\lambda})~.
\end{equation}
To make this more precise, let us analyze the conditions on the 3-Lie algebra-valued vector $C^\mu$ in further detail. Assuming the map $D:\CA\wedge\CA\rightarrow \frg_\CA$ is nondegenerate,\footnote{This condition holds, \eg, in the most interesting case of the 3-Lie algebra $A_4$.} then the equation $D(C^\mu,C^\nu)=0$ implies that $C^\mu$ can be factorized into its vector and its 3-Lie algebra part. Considering loop space, we have a natural candidate for this factorization,
\begin{equation}\label{eq:factorization}
 C^\mu=C \xd^\mu(\tau)~,
\end{equation}
where $C$ is a constant element of the 3-Lie algebra $\CA$ and $\xd^\mu(\tau)$ denotes the tangent vector to a loop $x^\mu(\tau)$. This decomposition of $C^\mu$ renders \eqref{eq:eomHF} indeed close to a transgression, as we will further discuss  below. First, however, let us consider the other constraints on $C^\mu$. We have
\begin{equation}
\begin{aligned}\label{eq:nabla}
 0=\nabla_\mu C^\nu=\dpar_\mu \xd^\nu(\tau)C+\xd^\nu (A_\mu\acton C):=&\left(\oint_{S^1}\dd \sigma\,\frac{\delta}{\delta x^\mu(\sigma)}\xd^\nu(\tau)\right)C+\xd^\nu (A_\mu\acton C)\\
=&\xd^\nu (A_\mu\acton C)~,
\end{aligned}
\end{equation}
where $\dpar_\mu$ is a shorthand notation for the loop space derivative
\begin{equation}
 \dpar_\mu:=\oint_{S^1}\dd \tau\,\frac{\delta}{\delta x^\mu(\tau)}~.
\end{equation}
Recall that the gauge field strength $F_{\mu\nu}$ and thus the gauge potential $A_\mu$ are elements of the subalgebra $\frg^C_\CA$ of the associated Lie algebra $\frg_\CA$ defined as
\begin{equation}
 \frg^C_{\CA}:=\span\{D(C,a)|a\in\CA\}~.
\end{equation}
We conclude that assuming \eqref{eq:eomHF} and \eqref{eq:factorization}, $\nabla_\mu C^\nu=0$ is trivially satisfied, as $[C,\dotsp,C]=0$. The remaining conditions
\begin{equation}
D(C,\xd^\rho\nabla_\rho X^I)=D(C,\xd^\rho\nabla_\rho\Psi)=D(C,\xd^\rho\nabla_\rho H_{\mu\nu\lambda})=0~,
\end{equation}
however, still have to be imposed.

Note that the components of all the 3-Lie algebra-valued fields $X^I$, $\Psi$ and $H_{\mu\nu\kappa}$ parallel to $C$ decouple from the equations of motion \eqref{eq:eomTensor}, as their interaction terms vanish. 

\subsection{Transgression of the 3-Lie algebra (2,0) tensor multiplet equations}

 We can now interpret equation \eqref{eq:eomHF} as the transgression-like map\footnote{Recall that we restrict ourselves to loops allowing for a parametrization such that
$\xd^\mu(\tau)\xd_\mu(\tau)=R^2$.}
\begin{equation}\label{eq:transgressionCondition}
 \cF_{\mu\nu}(x)=D\big(C^\lambda,H_{\mu\nu\lambda}(x(\tau))\big)=D\big(C,H_{\mu\nu\lambda}(x(\tau))\,\xd^\lambda(\tau)\big)~.
\end{equation}
We always use a circle to label fields on loop space. As mentioned under \eqref{eq:full}, we drop the integrals over angles in our equations to simplify notation. This is possible, as equation \eqref{eq:transgressionCondition} is in one-to-one correspondence with the full transgression-like map\footnote{Note the difference to the formulas in \cite{Huang:2010db}, where $F_{\mu\nu}(x):=\oint_{S^1} \dd \tau\, D(C,H_{\mu\nu\lambda}(x(\tau))\xd^\lambda(\tau))$.}
\begin{equation}
\begin{aligned}
 \cF(x)\big(v_1(x),v_2(x)\big)\ =\ &\cF_{\mu\nu}(x)\,v^\mu_1(x)\,v^\nu_2(x)\\
\ :=\ &\oint_{S^1} \dd \tau\, D(C,H_{\mu\nu\lambda}(x(\tau))\xd^\lambda(\tau))\,v^\mu_1(\tau)\,v^\nu_2(\tau)~,
\end{aligned}
\end{equation}
where $\cF$ is here seen as a map $T \CL M \wedge T\CL M\rightarrow \frg^C_\CA$. 

The additional map $D(C,\dotsp)$ in \eqref{eq:transgressionCondition} is evidently necessary to turn the 3-Lie algebra-valued 3-form $H$ into a gauge field strength taking values in the inner derivations $\frg^C_\CA$.

We find that the field $H$ can in fact be regarded as redundant on loop space, as all information is encoded in $\cF$. The components of $H$ which are in the kernel of $D(C,\dotsp)$ are flat gauge fields, which decouple from the equations of motion \eqref{eq:eomTensor} and can be set to zero up to a gauge transformation. The reinterpretation of the 3-form $H$ as a field strength on loop space is further motivated by the gauge transformation of $H$ postulated in \cite{Lambert:2010wm}
\begin{equation}
 \delta H_{\mu\nu\kappa}=\Lambda \acton H_{\mu\nu\kappa}=\Lambda^{ab}[\lambda_a,\lambda_b,H_{\mu\nu\kappa}]~,
\end{equation}
which is not the expected\footnote{In the abelian case, the gauge parameter carries an index: $\delta B_{\mu\nu}=\dpar_\mu\Lambda_\nu-\dpar_\nu\Lambda_\mu.$} gauge transformation law for a 3-form field strength. Moreover, as pointed out in \cite{Lambert:2010wm}, it does not seem possible to interpret $H$ as originating from a 2-form potential $B$ in a way compatible with the transgression-like map \eqref{eq:transgressionCondition}. In the following, we therefore aim at eliminating $H$ from both the equations of motion as well as the supersymmetry transformations to obtain a set of gauge equations on loop space.

The spinor field forms a nontrivial representation of the Poincar{\'e} group and should therefore also be transgressed. This is necessary for moving the supersymmetry transformations over to loop space later and also implies that we need to transgress the scalar fields. That is, we define\footnote{Our spinor conventions are the ones found in the appendix of \cite{Lambert:2010wm}.}
\begin{equation}\label{eq:3LAtransgression}
 \cX^I(x(\tau)):=R~D(C,X^I(x(\tau)))\eand\cPsi(x(\tau)):=\Gamma^\rho \xd_\rho D(C,\Psi(x(\tau)))~.
\end{equation}
Note that these field redefinitions have a nontrivial kernel: All components of $X^I$ and $\Psi$ along $C\in\CA$ are lost in going to $\cX$ and $\cPsi$. However, as was remarked in \cite{Lambert:2010wm}, these components furnish a free (2,0) tensor multiplet and as a result can be dealt with separately using the usual $\sU(1)$ transgression. The gauge potential is trivially lifted to the loop space $\CL\FR^{1,5}$.

With \eqref{eq:transgressionCondition} and \eqref{eq:3LAtransgression}, the tensor multiplet equations \eqref{eq:eomTensor} now reduce to equations which resemble the equations of motion of SYM theory with gauge algebra $\frg^C_\CA$. The Dirac equation, 
\begin{equation}
 \Gamma^\mu\nabla_\mu\Psi-[X^I,C^\nu,\Gamma_\nu\Gamma^I\Psi]=0~,
\end{equation}
for example, is reformulated according to
\begin{equation}
 \begin{aligned}
  D(C,\Gamma^\mu\nabla_\mu\Psi)-D(C,[X^I,C,\xd^\nu\Gamma_\nu\Gamma^I\Psi])&=0~,\\
  [C,\Gamma^\mu\nabla_\mu\Psi,a]+[C,[X^I,C,\Gamma^I\xd^\nu\Gamma_\nu\Psi],a]&=0~,\\
  \Gamma^\mu\nabla_\mu[C,\Psi,a]-[C,\Gamma^\mu\Psi,\nabla_\mu a]+\hspace{4.5cm}&\\
[X^I,C,[C,\Gamma^I\xd^\nu\Gamma_\nu\Psi,a]]-[C,\Gamma^I\xd^\nu\Gamma_\nu\Psi,[X^I,C,a]]&=0~,\\
  \tfrac{1}{R^2}(\Gamma^{\mu\nu}\xd_\nu\nabla_\mu\cPsi)\acton a-\tfrac{1}{R}(\Gamma^I [\cX^I,\cPsi])\acton a&=0~, 
 \end{aligned}
\end{equation}
where $a\in\CA~$ and we used that $\xd^\mu\Gamma_\mu\xd^\nu\Gamma_\nu=\frac{1}{2}\xd^\mu\xd^\nu\{\Gamma_\mu,\Gamma_\nu\}=R^2$. This yields
\begin{equation}
  \tfrac{1}{R} \Gamma^{\mu\nu}\xd_\nu\nabla_\mu\cPsi-\Gamma^I [\cX^I,\cPsi]=0~.
\end{equation}
The equation of motion for the scalar field is correspondingly rewritten as
\begin{equation}
 \nabla^2 \cX^I+ \tfrac{\di}{2} \tfrac{1}{R}\xd^\nu[\bar{\cPsi},\Gamma_\nu\Gamma^I\cPsi]- [\cX^J,[\cX^J,\cX^I]]=0~.
\end{equation}
Because the 3-form field $H$ is selfdual, we obtain two equations of motion from \eqref{eq:eomTensor} after the rewriting \eqref{eq:transgressionCondition}. First, we have 
\begin{equation}\label{eq:Bianchi}
 D\big(C\xd^\lambda,\nabla_{[\mu}H_{\nu\kappa\lambda]}+\tfrac{1}{4}\eps_{\mu\nu\kappa\lambda\sigma\tau}[X^I,\nabla^\tau X^I,C \xd^\sigma]+\tfrac{\di}{8}\eps_{\mu\nu\kappa\lambda\sigma\tau}[\bar{\hat{\Psi}},\Gamma^\tau\hat{\Psi},C\xd^\sigma]\big)=\nabla_{[\mu}\cF_{\nu\kappa]}=0~,
\end{equation}
where we used $D(C\xd^\lambda,\nabla_\lambda H_{\mu\nu\kappa})=0$ and $\xd^{[\rho}\xd^{\sigma]}=0$. Equation \eqref{eq:Bianchi} is simply the Bianchi identity for the field strength $\cF$. Substituting $H$ by $*H$ in \eqref{eq:eomTensor}, we also find the rewritten equation
\begin{equation}
\nabla_\mu \cF^{\mu\nu}+[\cX^I,\nabla^\nu \cX^I]+ \di\left( [\bar{\cPsi},\Gamma^\nu\cPsi] - \tfrac{1}{R^2}\xd^\sigma \xd^\nu[\bar \cPsi, \Gamma_\sigma \cPsi]\right)=0~. 
\end{equation}
Note that the selfduality condition on $H$ disappeared in the loop space picture. The supersymmetry transformations now read as
\begin{equation}
\begin{aligned}
 \delta \cX^I&= \tfrac{1}{R} \di \epsb \Gamma^I\xd^\rho\Gamma_\rho\cPsi~,\\
 \delta \cA_{\mu}&=\tfrac{1}{R^2} \di\epsb \Gamma_{\mu\lambda}\Gamma_\rho\xd^\lambda\xd^\rho\cPsi~,\\
  \delta \cPsi&=\tfrac{1}{R}\Gamma^{\nu\mu}\xd_\nu\Gamma^I\nabla_\mu\cX^I\eps+\tfrac{1}{2}\Gamma_{\mu\nu}\cF^{\mu\nu}\eps-\tfrac{1}{2}\Gamma^{IJ}[\cX^I,\cX^J]\eps~,
\end{aligned}
\end{equation}
where we have made use of the identities
\begin{equation}
\begin{aligned}
 \Gamma_\sigma\Gamma_{\mu\nu\kappa}= \Gamma_{\sigma\mu\nu\kappa} +3\eta_{\sigma[\mu}\Gamma_{\nu\kappa]}~,~\\
\Gamma_{\mu\nu\kappa\lambda}=\tfrac{1}{2}\eps_{\mu\nu\kappa\lambda\rho\sigma}\Gamma^{\rho\sigma}\Gamma_{012345}~.
\end{aligned}
\end{equation}

\subsection{Supersymmetric gauge field equations on loop space}

Let us summarize our results. We showed that the equations \eqref{eq:eomTensor} found in \cite{Lambert:2010wm} correspond to supersymmetric Yang-Mills-like equations on loop space with the matter fields \eqref{eq:3LAtransgression}. The extended transgression involves not only lowering the degree of forms by one (through contraction with the vector tangent to the loop for $H$ and $\Psi$) but also mapping from the 3-Lie algebra $\mathcal A$ to the Lie algebra $\frg^C_\CA$ through the map $D(C,\dotsp)$ for all fields.  

Hence, by rewriting the 3-Lie algebra (2,0) tensor multiplet equations in loop space, we have found that the supersymmetry algebra associated with the transformations
\begin{equation}\label{eq:susysummary}
\begin{aligned}
 \delta \cX^I&=\tfrac{1}{R}\di \epsb \Gamma^I\xd^\rho\Gamma_\rho\cPsi~,\\
 \delta \cA_{\mu}&=\tfrac{1}{R^2} \di\epsb \Gamma_{\mu\lambda}\Gamma_\rho\xd^\lambda\xd^\rho\cPsi~,\\
  \delta \cPsi&=\tfrac{1}{R}\Gamma^{\nu\mu}\xd_\nu\Gamma^I\nabla_\mu\cX^I\eps+\tfrac{1}{2}\Gamma_{\mu\nu}\cF^{\mu\nu}\eps-\tfrac{1}{2}\Gamma^{IJ}[\cX^I,\cX^J]\eps~,
\end{aligned}
\end{equation}
closes up to the equations of motion
\begin{equation}\label{eq:eomsummary}
\begin{aligned}
 \nabla^2 \cX^I+\tfrac{\di}{2}\tfrac{1}{R}\xd^\nu[\bar{\cPsi},\Gamma_\nu\Gamma^I\cPsi]+ [\cX^J,[\cX^J,\cX^I]]&=0~,\\
\tfrac{1}{R} \Gamma^{\mu\nu}\xd_\nu\nabla_\mu\cPsi-\Gamma^I [\cX^I,\cPsi]&=0~,\\
\nabla_\mu \cF^{\mu\nu}+[\cX^I,\nabla^\nu \cX^I]+\di\left( [\bar{\cPsi},\Gamma^\nu\cPsi] - \tfrac{1}{R^2}\xd^\sigma \xd^\nu[\bar \cPsi, \Gamma_\sigma \cPsi]\right)&=0~,
\end{aligned}
\end{equation}
along with the Bianchi identity $\nabla_{[\mu}\cF_{\nu\lambda]}=0$ and the constraints
\begin{equation}\label{eq:constraintssummary}
\xd^\mu \nabla_\mu \cX^I = \xd^\mu \nabla_\mu \cPsi  = \xd^\mu \nabla_\mu \cF_{\nu\lambda} =0~.
\end{equation}
Note that the tangent vector of the loop cannot be eliminated from the equations.

Above, we started from fields taking values in a 3-Lie algebra $\CA$ with associated Lie algebra $\frg_\CA$. We then mapped these fields to ones on loop space taking values in the restriction $\frg^C_\CA$ of $\frg_\CA$. Note that using the so-called Lorentzian 3-Lie algebras \cite{Gomis:2008uv,Benvenuti:2008bt,Ho:2008ei} and direct sums thereof, we can construct arbitrary such restrictions $\frg^C_\CA$, see \eg\ \cite{DeMedeiros:2008zm}. This implies that equations \eqref{eq:susysummary}-\eqref{eq:constraintssummary} can involve fields taking values in an arbitrary gauge algebra. 

Therefore, these equations describe a new set of supersymmetric field equations for the $\CN=2$ vector supermultiplet in five dimensions lifted to loop space. The constraints are necessary in order to obtain the correct counting of physical degrees of freedom in the supersymmetry multiplet: They reduce the four gauge degrees of freedom in six dimensions down to the three of a selfdual 3-form field strength via $\xd^\rho\nabla_\rho \cF_{\mu\nu}=0$.\footnote{It is not entirely clear what the meaning of a supersymmetry multiplet would be on the infinite-dimensional loop space $\CL\FR^{1,5}$. Na\"ively, one expects a one-parameter family of supermultiplets on $\FR^{1,5}$ with the parameter being the loop angle.}

One might be tempted to try to write down an action functional for equations \eqref{eq:eomsummary} and \eqref{eq:constraintssummary}. For this, one would assume that the constraints \eqref{eq:constraintssummary} arise from varying the action with respect to some Lagrange multipliers. However, the integrability condition necessary for the existence of an action functional is violated by \eqref{eq:eomsummary} and \eqref{eq:constraintssummary}.

\subsection{Reduction to five-dimensional super Yang-Mills theory}

Having recast all of our equations in terms of expressions on loop space, it is straightforward to see the reduction to five-dimensional SYM theory. We start by picking a particular direction, say $x^5$, which is to be interpreted as the M-theory direction. Following the standard recipe, we then have to compactify this direction and identify its radius with the square of the Yang-Mills coupling. That is, we turn the loop space $\CL \FR^{1,5}$ into the loop space of $\FR^{1,4}\times S^1$ and define the radius of the contained $S^1$ to be $R=g_{\rm YM,5D}^2$. The reduction is now performed by simply restricting ourselves to loops wrapping the M-theory direction, so that  $x^\mu(\tau) = R\delta_5^\mu \tau$. Then $\xd^\mu = R\delta^\mu_5$ and due to the constraints $\xd^\rho \nabla_\rho \cX^I=\xd^\rho \nabla_\rho \cPsi=\xd^\rho \nabla_\rho \cF_{\mu\nu}=0$, all dependence on the compactified direction, and hence the loop parameter is eliminated. Moreover, the six-dimensional gauge field strength $\cF_{\mu\nu}$ is reduced to a field strength in five dimensions. Identifying further $\cPsi = \Gamma_5 \Psi_{\rm YM}$, the equations of motion \eqref{eq:eomsummary} as well as the supersymmetry transformations \eqref{eq:susysummary} reduce precisely as expected. We thus recover five-dimensional maximally supersymmetric Yang-Mills theory.

When the compactified direction is small, the loop practically vanishes and one has a local theory. When the loop is very large, the SYM theory is very strongly coupled. In our case, the constraints \eqref{eq:constraintssummary} enforce that no field carries momentum along the $x^5$ direction for any value of $R$, so there is no momentum in the M-theory direction. It has been argued \cite{Douglas:2010iu,Lambert:2010iw} that instanton-particles of five-dimensional SYM theory precisely capture all the degrees of freedom associated with the Kaluza-Klein (KK) modes of the reduction of the $(2,0)$ theory on $S^1$. As a result, SYM theory at infinite coupling can in turn be used to define the $(2,0)$ theory. It would be interesting to relate this infinite tower of instantons (or KK modes) to the infinite numbers of fields that enter the nonlocal description in terms of unconstrained fields on loop space.

\section{Nonabelian selfdual strings on loop space}\label{nonabeliansds}

We now move on to demonstrating that our reinterpretation is not only consistent but also allows for an interesting application: We can extend a recently found construction of selfdual strings \cite{Saemann:2010cp} from the abelian to the nonabelian case. 

\subsection{BPS equations and selfdual strings}

We are interested in BPS equations which can be interpreted in terms of stacks of M2-branes ending on stacks of M5-branes in the following way:
\begin{equation}\label{diag:Branes}
\begin{aligned}
\begin{tabular}{rccccccccccc}
 & 0 & 1 & 2 & 3 & \phantom{(}4\phantom{)} & 5 & 6 & 7 & 8 & 9 & 10\\
M2 & $\times$ & & & & & $\times$ & $\times$ \\
M5 & $\times$ & $\times$ & $\times$ & $\times$ & $\times$ & $\times$ &
\end{tabular}
\end{aligned}
\end{equation}
We are therefore looking for BPS solutions to the equations \eqref{eq:eomTensor}, for which $\Phi:=X^6\neq 0=X^7,\ldots,X^{10}$ and which preserves half of the supersymmetry. This condition, along with $H_{0ij} = H_{5ij} = 0$ and the constraint $\nabla_5 \Phi = 0$,  reduce $\delta\Psi=0$ to
\begin{equation}\label{eq:BPS}
 \Gamma^i\Gamma^6\nabla_i \Phi\eps+\tfrac{1}{2\times 3!}\Gamma_{ijk}H^{ijk}\eps=0~,~~~i,j,k=1,\ldots,4~.
\end{equation}
By imposing $\Gamma^{05}\eps=\Gamma^6\eps$ we break  half of the supersymmetry, as expected for having M2-branes extending in the $x^0,x^5,x^6$ directions. Hence, \eqref{eq:BPS} simplifies further to
\begin{equation}\label{eq:sdseqn}
 H_{ijk}=\eps_{ijkl}\nabla^l \Phi~.
\end{equation}
In the abelian case, this reduces to the equation describing a selfdual string, cf.\ \cite{Howe:1997ue}
\begin{equation}
 H_{05i}=-\dpar_i\Phi\eand H_{ijk}=\eps_{ijkl}\dpar_l\Phi~.
\end{equation}
The transgression of the second equation to the free loop space of $\FR^4$ reads as 
\begin{equation}\label{eq:LoopSpaceSDS}
 \cF_{ij}\big(x(\tau)\big)=\eps_{ijkl}\frac{\xd^k(\tau)}{R} \der{x^l}\cPhi\big(x(\tau)\big)~,
\end{equation}
and the latter was used in the lift of the ADHMN construction from monopoles to selfdual strings \cite{Saemann:2010cp}.\footnote{Note that the additional factor of $\frac{1}{R}$ is due to the fact that in \cite{Saemann:2010cp}, the Higgs field $\cPhi$ was not rescaled by $R$ when going to loop space, cf.\ equation \eqref{eq:3LAtransgression}.} It should be interpreted as describing a configuration \eqref{diag:Branes} with multiple M2-branes ending on a single M5-brane.

The transgression of \eqref{eq:sdseqn} in the nonabelian case to fields living in a subalgebra of the associated Lie algebra $\frg_\CA$ was already suggested in \cite{Saemann:2010cp} and reads as 
\begin{equation}\label{eq:sdstrans}
 \cF_{ij}=\eps_{ijkl}\frac{\xd^k}{R}\nabla^l \cPhi~.
\end{equation}
Here, it appears naturally from the nonabelian tensor multiplet equations \eqref{eq:eomTensor}. Note that solutions to \eqref{eq:sdstrans} automatically solve the full equations \eqref{eq:eomsummary}, since
\begin{equation}
 \nabla^i \cF_{ij}=\eps_{ijkl}\frac{\xd^k}{R}\nabla^i\nabla^l \cPhi=-\tfrac{1}{2}\eps_{ijkl}\frac{\xd^k}{R}[\cF^{il},\cPhi]=[\nabla_j \cPhi,\cPhi]~,
\end{equation}
and we have used the constraint $\xd^i \nabla_i \Phi = 0$. Equation \eqref{eq:sdstrans} could thus potentially describe the effective dynamics of configuration \eqref{diag:Branes}, with a stack of M2-branes ending on multiple M5-branes.

Recall that for a Dirac monopole, the magnitude of the Higgs field diverges at the location of the monopole. One therefore describes a Dirac monopole usually in terms of a principal $\sU(1)$-bundle over a sphere $S^2$ with the monopole at its center. Something similar was observed for the abelian selfdual string: The magnitude of the Higgs field diverges and one has to describe it by a local abelian gerbe over $S^3$. This is why the transgression used in \cite{Saemann:2010cp} led to the loop space of $S^3$. This space $\CL S^3$ was described in terms of loops $x^i(\tau)$ satisfying the conditions
\begin{equation}\label{eq:LSrelations}
 x^i(\tau)x^i(\tau)=R^2~\Rightarrow~x^i(\tau)\xd^i(\tau)=0\eand \xd^i(\tau)\xd^i(\tau)=R^2~,
\end{equation}
where the last equation is identical to our choice of parametrization of the loops.

For monopoles in Yang-Mills theory with gauge group $\sSU(n)$, $n\geq 2$, the Higgs field does not have to diverge, and one can switch to a description in terms of a principal $\sSU(n)$-bundle over $\FR^3$. The same is expected to happen for nonabelian selfdual strings, and we are able to use the loop space of $\FR^4$.

\subsection{The abelian case}

We now briefly review the construction of selfdual strings in the abelian case as developed originally in \cite{Saemann:2010cp}. Underlying this construction is the duality between the M5-brane and M2-brane perspectives on the configuration \eqref{diag:Branes} for a single M5-brane. Recall that the original ADHMN construction \cite{Nahm:1979yw,Nahm:1981nb,Nahm:1982jb} provides a transition between solutions of the Nahm equation and those of the Bogomolny monopole equation. The procedure of \cite{Saemann:2010cp} provides a similar link between solutions to the Basu-Harvey equation and solutions to the selfdual string equation in its transgressed form \eqref{eq:LoopSpaceSDS}.

Besides the 3-Lie algebra $\CA$ which appears in equations \eqref{eq:eomTensor} and is associated with the six-dimensional theory, consider a second, metric 3-Lie algebra $\CB$, which is used in the Basu-Harvey equation \cite{Basu:2004ed} in the three-dimensional context. The latter equation lives on an open interval $\CI\subset\FR$, which corresponds to one of the worldvolume directions of a stack of M2-branes suspended between M5-branes as in \eqref{diag:Branes}. In the case of a single M5-brane, we have $\CI=\FR^+$. The Basu-Harvey equation describes the dynamics of four scalar fields $X^i(s)$, $s\in \CI$, capturing the transverse fluctuations of the M2-branes. In 3-Lie algebra form, it reads as 
\begin{equation}\label{eq:BH}
 \dder{s} X^i=\tfrac{1}{3!}\eps^{ijkl}[X^j,X^k,X^l]~,~~~X^i(s)\in\CB~.
\end{equation}
Given a solution $X^i(s)$ to this equation, we construct a twisted Dirac operator
\begin{equation}
 \nablas_{s,x}=-\gamma_5\dder{s}+\gamma^{ij}\left(\tfrac{1}{2}D(X^i,X^j)-\di x^i(\tau)\xd^j(\tau)\right)~,
\end{equation}
where $x^i(\tau)$ describes an element of the loop space $\CL S^3$ of $S^3$ for reasons explained above. Moreover, $\gamma^{ij}=\tfrac{1}{2}[\gamma^i,\gamma^j]$, where the $\gamma^i$ are generators of $\sSpin(4)$ satisfying $\{\gamma^i,\gamma^j\}=2\delta^{ij}$. The form of this Dirac operator can be justified by symmetry considerations and by lifting a D1-D3-brane configuration described by the Nahm equation to M-theory \cite{Saemann:2010cp}. Consider now a zero mode $\psi_{s,x}$ of the adjoint of the twisted Dirac operator $\nablabs_{s,x}$. Such a zero mode is a ``spinor'' $\psi_{s,x}\in\FC^4\otimes \CB\otimes W^{1,2}(\CI)$, where $W^{1,2}(\CI)$ is the Sobolev space of functions on $\CI$ which are square integrable up to their first derivative. We normalize the zero mode according to 
\begin{equation}\label{eq:normalizePsiM}
 1=\int_\CI\dd s\, \psib_{s,x}\psi_{s,x}~.
\end{equation}
Analogously to the ADHMN procedure, we can construct a gauge potential and a Higgs field on loop space from this normalized zero mode
\begin{equation}\label{eq:MFieldDefinitions}
 \cA_i\big(x(\tau)\big)=\int_\CI \dd s\, \psib_{s,x} \der{x^i} \psi_{s,x}\eand\cPhi\big(x(\tau)\big)=-\di R\int_\CI \dd s\, \psib_{s,x}\,s\,\psi_{s,x}~.
\end{equation}
As one can show by explicit computation, these fields satisfy the loop space selfdual string equation \eqref{eq:LoopSpaceSDS}, cf.\ \cite{Saemann:2010cp}. The corresponding calculation in the nonabelian case is presented below.

\subsection{Nonabelian extension}

We now turn to the extension of \eqref{eq:MFieldDefinitions}, which yields a construction of solutions to the equations \eqref{eq:sdstrans} for gauge group $\sU(n)$. This might be related to allowing for multiple M5-branes in configuration \eqref{diag:Branes}. We thus want to start from a solution to the Basu-Harvey equation, from which we derive a suitable gauge potential $\cA$ and a scalar field $\cPhi$, both living in $\au(n)$.

The Dirac operator is the same as in the abelian case. The only differences are that we now work with the loop space $\CL\FR^4$ and that we consider multiple zero modes living in $\psi\in\FC^4\otimes \FC^n\otimes W^{1,2}(\CI)\otimes \CB$:
\begin{equation}
 \psi_{s,x}=\psi^{A,a}_x(s) \kappa_A\otimes e_a~,
\end{equation}
where $(\kappa_A)$ are the generators of the metric 3-Lie algebra $\CB$ and the $e_a$, $a=1,\ldots,n$, form a basis of $\FC^n$. The zero modes are normalized according to 
\begin{equation}\label{eq:normalizePsiMNA}
 \int_\CI\dd s\, (\psib^a_{s,x},\psi^b_{s,x})=\delta^{ab}~.
\end{equation}
Here, $(\dotsp,\dotsp)$ denotes the inner product on $\CB$. The Basu-Harvey equation appears in this construction as the condition that the operator $\Delta_{s,x}=\nablabs_{s,x}\nablas_{s,x}$ is invertible and commutes with $\gamma^5$ and $\gamma^{ij}$. There is thus a Green's function $G_x(s,t)$ satisfying $\Delta_{s,x}G_x(s,t)=-\delta(s-t)$, and we arrive at the projector
\begin{equation}
 \CP_x(s,t)=-\nablas_{s,x}G_x(s,t)\nablabs_{t,x}=\delta(s-t)-\psi_{s,x}^a(\psib_{t,x}^a,\dotsp)~.
\end{equation}
The nonabelian loop space selfdual string equation \eqref{eq:sdstrans} is now solved by the following fields
\begin{equation}
 (\cA_i)^{ab}=\int_\CI \dd s\,(\psib^a_{s,x},\dpar_i\psi^b_{s,x})\eand \cPhi^{ab}=-\di R\int_\CI \dd s\,(\psib^a_{s,x},s\psi^b_{s,x})~,
\end{equation}
as one sees by going through the explicit computation
\begin{equation*}
 \begin{aligned}
  (\cF_{ij})^{ab}&=2\int_\CI \dd s\,\big(\dpar_{[i}\psib^a_{s,x},\dpar_{j]}\psi^b_{s,x}\big)+2\int_\CI \dd s\int_\CI \dd t\,\big(\psib^a_{s,x},\dpar_{[i}\psi^c_{s,x}\big)\big(\psib^c_{s,x},\dpar_{j]}\psi^b_{s,x}\big)\\
&=-2\int_\CI \dd s\int_\CI \dd t\,\Big(\dpar_{[i}\psib^a_{s,x}\,,\,\left(\nablas_{s,x}G_x(s,t)\nablabs_{t,x}\right)\dpar_{j]}\psi^b_{t,x}\Big)\\
&=\int_\CI \dd s\int_\CI \dd t\,\Big(\psib_{s,x}^a,\left(\gamma_{ik}\xd^k G_x(s,t)\gamma_{jl}\xd^l-\gamma_{jk}\xd^k G_x(s,t)\gamma_{il}\xd^l\right)\psi^b_{t,x}\Big)~.
 \end{aligned}
\end{equation*}
Recall that the Green's function commutes with the $\gamma^{ij}$ and $\gamma_5$. Together with the identity
\begin{equation}
 [\gamma^{ik},\gamma^{jl}]\xd^k\xd^l=-2\eps_{ijmn}\gamma^{nk}\gamma_5\xd^m\xd^k~,
\end{equation}
we thus arrive at
\begin{equation}
 \begin{aligned}
(\cF_{ij})^{ab}&=-\eps_{ijmn}\int_\CI \dd s\int_\CI \dd t\,\Big(\psib_{s,x}^a,\,\left(2\gamma^{nk}\gamma_5G_x(s,t)\xd^m\xd^k\right)\psi^b_{t,x}\Big)\\
&=-\di\eps_{ijmn}\frac{\xd^m}{R}R\int_\CI \dd s\,\Big(\nabla_n\psib^a_{s,x},\,s\,\psi^b_{s,x}\Big)+\Big(\psib^a_{s,x},\,s\,\nabla_n\psi^b_{s,x}\Big)\\
&=\eps_{ijkl}\frac{\xd^k}{R}\nabla_l\cPhi^{ab}~.
 \end{aligned}
\end{equation}

In order for this to be a true BPS solution of the nonabelian selfdual string equations on loop space, the fields have to satisfy the additional constraints
\begin{equation}\label{eq:BPSconstraints}
 \xd^i\nabla_i \cPhi=\xd^i\nabla_i \cF=0~.
\end{equation}
Choosing the gauge $\xd^\rho A_\rho=0$, this condition is satisfied by all fields depending on the loops through $x^{[i}\xd^{j]}$, as $\xd^k\dpar_k(x^{[i}\xd^{j]})=0$. The form of our Dirac operator now induces exactly this functional dependence onto the zero modes and thus onto the fields. We therefore expect the constructed solutions to satisfy the constraints \eqref{eq:BPSconstraints} automatically.

\subsection{Explicit solution for `one M2-brane between two M5-branes'}

As an example for our construction, let us briefly discuss the simplest possible nonabelian case. Underlying this example is configuration \eqref{diag:Branes} with a single `M2-brane' between two `M5-branes' located at the endpoints of the interval $\CI=(-s_0,s_0)$. 

On the six-dimensional side, we expect the fields to live in a subalgebra $\frg_\CA^C$ of the associated Lie algebra $\frg_\CA=\asu(2)\oplus\asu(2)$ of the 3-Lie algebra $\CA=A_4$. We choose generators $\gamma^{ij}\gamma_5$ for $\asu(2)\oplus\asu(2)$ and in Weyl representation, these generators are block diagonal. Restricting to $\frg_\CA^C$ here corresponds to restricting to the diagonal of $\asu(2)\oplus\asu(2)$. For simplicity, we will allow for all fields to live in $\frg_\CA$. The restriction to $\frg_\CA^C$ can be trivially performed in the end. 

On the three-dimensional side, the 3-Lie algebra $\CB$ is abelian, and the Basu-Harvey equation turns into $\dder{s}X^\mu=0$. Thus, we can choose $X^\mu=0$, which leaves us with the adjoint of the Dirac operator
\begin{equation}
 \nablabs_{s,x}=\gamma_5\dder{s}+\gamma^{ij}\di x^i(\tau)\xd^j(\tau)~.
\end{equation}
The normalized zero modes of this operator, arranged in matrix form, read as
\begin{equation}
 \psi_{s,x}=\sqrt{\frac{\beta}{\sinh(2 \beta s_0)}}\left(\cosh(\beta s)\unit_4+\frac{\di}{\beta}\sinh(\beta s)x^i\xd^j\gamma^{ij}\gamma_5\right)~,~~~\beta^2=2(x^{[i}\xd^{j]})^2~.
\end{equation}
From the zero modes, one derives the Higgs field
\begin{equation}
 \mathring{\Phi}\big(x(\tau)\big)=\frac{\di R}{2 \beta^2}\big(2\beta s_0\coth(2 \beta s_0)-1\big)x^i\xd^j\gamma^{ij}\gamma_5~,
\end{equation}
as well as a more complicated looking gauge potential, both of which indeed satisfy the nonabelian loop space selfdual string equation \eqref{eq:sdstrans}. This solution is very similar to the corresponding solution of a charge one monopole in $\sSU(2)$ Yang-Mills theory, which reads as
\begin{equation}
 \Phi_{\rm monopole}=\frac{\vec{x}\cdot\vec{\sigma}}{|\vec{x}|^2}\big(s_0 |\vec{x}|\coth(s_0 |\vec{x}|)-1\big)~.
\end{equation}
We will say more about the reduction in the next section. Note that as expected, our solution indeed satisfies the constraint $\xd^i\dpar_i \Phi=0$, as it only depends on the loop $x(\tau)$ through the product $x^{[i}\xd^{j]}$. 

\subsection{Reduction to monopoles}

The above results can be readily reduced to monopoles in nonabelian Yang-Mills theory. This is done simultaneously in the loop space picture as well as at the level of the Basu-Harvey equation. As we have already discussed, the reduction on loop space is performed by following a procedure similar to that of the full theory, cf.\ \cite{Saemann:2010cp} for the abelian case. That is, we replace the loop space $\CL \FR^{4}$ by the loop space of $\FR^{3}\times S^1$ and restrict ourselves to loops around the compactified direction: $x^\mu(\tau)=R\tau\delta^\mu_4$, $\xd^\mu(\tau)=R\delta^\mu_4$. We once again identify $R=g_{\rm YM,5D}^2$. As the coordinate $x^\mu(\tau)$ always appears in an antisymmetrized combination with $\xd^\mu(\tau)$, this implies that $x^4(\tau)$ drops out from all equations. Explicitly, we have
\begin{equation}
 \cF_{ij}=\eps_{ijkl}\frac{\xd^k}{R}\nabla^l \cPhi~~~\rightarrow~~~\cF_{\alpha\beta}=-\eps_{\alpha\beta\gamma}\nabla^\gamma \cPhi~,
\end{equation}
for $\alpha= 1,2,3$, which is the Bogomolny monopole equation. In the Dirac operator, the generators $\gamma^{ij}$ of $\sSpin(4)$ are reduced to $\gamma^{4\alpha}$, generating $\sSU(2)\cong\sSpin(3)\subset \sSpin(4)$. 

Additionally, we have to restrict the 3-Lie algebra $\CB$ appearing in the Basu-Harvey equation to a Lie algebra. This is done analogously to \cite{Mukhi:2008ux}: If $X^4$ has a vacuum expectation value $\langle X^4\rangle= v $, we can perform an expansion for large $v$. The Basu-Harvey equation then reduces to the Nahm equation plus subleading terms of order $\mathcal{O}(\frac{1}{v})$.\footnote{In fact, one could have performed this expansion in the full BLG Lagrangian, which reduces to three-dimensional SYM theory with coupling $g^2_{\rm YM,3D} = \frac{v^2}{k}$. Looking there for a BPS equation, one would also obtain the Nahm equation.} In total, our Dirac operator is reduced to that of the ordinary ADHMN-construction.

One might be alarmed by the fact that the reduction to the Nahm equation gives lower order corrections in the large-$v$ expansion, while the reduction of the nonabelian selfdual string equation does not contain such corrections. However, there is no discrepancy if one takes into account the regime in which the two sets of equations overlap: Solutions to \eqref{eq:BH} should describe a set of coincident branes expanding into a higher-dimensional brane configuration through a fuzzy funnel \cite{Constable:1999ac}, as one moves in the $x^6 \equiv s$ direction of the three-dimensional worldvolume. For the semi-infinite funnel, one finds that $\Phi \sim \frac{1}{\sqrt s}$, and this is in turn proportional to the radius of the fuzzy sphere cross-section in the transverse directions at each fixed $s$. Therefore, the fuzzy sphere is very small at $s=\infty$ and diverges at $s=0$, which signals the presence of the higher dimensional brane.\footnote{Recall that for this picture to make sense, one needs to take the large-$N$ limit, which can only happen in the full ABJM theory.} It is then clear that close to $s=0$, where the two descriptions should overlap, and for $v \sim \frac{1}{\sqrt s}$, the $\mathcal O (\frac{1}{v})$ corrections are negligible. Moreover, since  the dynamics in terms of our higher dimensional brane picture are always those of a nonabelian SYM theory in five dimensions, one immediately concludes that the fuzzy sphere involved is an $S^2$ rather than an $S^3$. This is an important consistency check and in agreement with the fluctuation analysis performed in \cite{Nastase:2009ny} for the ABJM theory around fuzzy funnel vacua, which explicitly displayed the $S^2$ geometry.

\section{Conclusions}\label{sec:conclusions}

In this paper, we found that the supersymmetric equations of motion for the 3-Lie algebra (2,0) tensor multiplet derived in \cite{Lambert:2010wm} admit a very natural interpretation in terms of gauge field equations on loop space. In this interpretation, the reduction procedure from the loop space equations to SYM theory follows the standard recipe of reducing M-theory to string theory. Irrespective of their applications to M-theory, we believe that the supersymmetric gauge field equations on loop space that we obtained are interesting in their own right: They provide a new supersymmetric set of equations for the field content of five-dimensional $\CN=2$ SYM theory lifted to loop space.

We also studied the BPS sector of the supersymmetric gauge field equations on loop space. These led to the nonabelian generalization of the selfdual string soliton proposed in \cite{Saemann:2010cp}. An ADHMN-like construction can be established, which yields such nonabelian BPS solutions in loop space from solutions to the Basu-Harvey equation. Using this procedure, we derived the simplest generalized selfdual string solution explicitly. Note that the Basu-Harvey equation considered here is the BPS equation of the BLG model based on the 3-Lie algebra $A_4$ with associated Lie group $\sSU(2) \times \sSU(2)$. This group has finite rank and therefore clearly does not capture the dynamics of stacks of $n$ M2-branes, unless $n=2$ and the Chern-Simons level is $k=1$. To describe arbitrarily many M2-branes, one should switch from the BLG model to the ABJM model. A paper studying the ADHMN-like construction for the generalized 3-Lie algebras of \cite{Bagger:2008se} leading to the ABJM model (as well as for those of \cite{Cherkis:2008qr}, yielding $\CN=2$ BLG-type theories) is in preparation \cite{Palmer:2011vx}. The structures remain mostly the same.

There are essentially two directions which deserve future study in our opinion. First, recall that the transgression to loop space was a tool that, in the abelian case, allowed us to avoid dealing with gerbes trough the transgression map. Assuming that our loop space equations provide the setup for describing nonabelian selfdual strings and certain aspects of stacks of multiple M5-branes, it would be worth trying to develop the corresponding descriptions in terms of nonabelian gerbes, which might lead to equations of motion for unconstrained fields.

Second, the fact that our equations become equivalent to five-dimensional SYM theory after reduction by putting $x^\mu(\tau)=\tau\delta^\mu_5$ is due to the constraints $\xd^\mu \nabla_\mu \cX^I = \xd^\mu \nabla_\mu\cPsi=\xd^\mu\nabla_\mu \cF_{\mu\nu}=0$. These  constraint are essential for the supersymmetry algebra to close. In the reduction, they truncate the infinite number of fields in the loop description down to the zero mode sector. Obviously, it would be very interesting to try and obtain a generalization of this supersymmetric system that does not need the constraints for closure. Moreover, if the higher modes lost due to the constraints are to correspond to Kaluza-Klein modes on the M-theory circle, then according to \cite{Douglas:2010iu,Lambert:2010iw} they also correspond to excitations that carry instanton charge in 5D. It would therefore be intriguing to examine to what extent a description in terms of unconstrained fields in loop space can capture these nonperturbative degrees of freedom and hence M-theory physics.

\section*{Acknowledgments}

We would like to thank David Berman for useful discussions. C.P.\ is supported by the STFC rolling grant ST/G000395/1. C.S.\ is supported by a Career Acceleration Fellowship from the UK Engineering and Physical Sciences Research Council.

\bibliographystyle{latexeu}

\bibliography{bigone}

\begin{thebibliography}{10}

\bibitem{Bagger:2006sk}
J.~Bagger and N.~Lambert,
{\em Modeling multiple M2's,}
\href{http://dx.doi.org/10.1103/PhysRevD.75.045020}{Phys. Rev. D {\bf 75}
  (2007) 045020} [{\tt
  \href{http://www.arxiv.org/abs/hep-th/0611108}{hep-th/0611108}}].

\bibitem{Gustavsson:2007vu}
A.~Gustavsson,
{\em Algebraic structures on parallel M2-branes,}
\href{http://dx.doi.org/10.1016/j.nuclphysb.2008.11.014}{Nucl. Phys. B {\bf
  811} (2009)~66} [{\tt \href{http://www.arxiv.org/abs/0709.1260}{0709.1260
  [hep-th]}}].

\bibitem{Bagger:2007jr}
J.~Bagger and N.~Lambert,
{\em Gauge symmetry and supersymmetry of multiple M2-branes,}
\href{http://dx.doi.org/10.1103/PhysRevD.77.065008}{Phys. Rev. D {\bf 77}
  (2008) 065008} [{\tt \href{http://www.arxiv.org/abs/0711.0955}{0711.0955
  [hep-th]}}].

\bibitem{Bagger:2007vi}
J.~Bagger and N.~Lambert,
{\em Comments on multiple M2-branes,}
\href{http://dx.doi.org/10.1088/1126-6708/2008/02/105}{JHEP {\bf 02} (2008)
  105} [{\tt \href{http://www.arxiv.org/abs/0712.3738}{0712.3738 [hep-th]}}].

\bibitem{Bagger:2008se}
J.~Bagger and N.~Lambert,
{\em {Three-algebras and $\CN=6$ Chern-Simons gauge theories},}
\href{http://dx.doi.org/10.1103/PhysRevD.79.025002}{Phys. Rev. D {\bf 79}
  (2009) 025002} [{\tt \href{http://www.arxiv.org/abs/0807.0163}{0807.0163
  [hep-th]}}].

\bibitem{Aharony:2008ug}
O.~Aharony, O.~Bergman, D.~L.~Jafferis, and J.~Maldacena,
{\em {$\CN=6$ superconformal Chern-Simons-matter theories, M2-branes and their
  gravity duals},}
\href{http://dx.doi.org/10.1088/1126-6708/2008/10/091}{JHEP {\bf 10} (2008)
  091} [{\tt \href{http://www.arxiv.org/abs/0806.1218}{0806.1218 [hep-th]}}].

\bibitem{Lambert:2010wm}
N.~Lambert and C.~Papageorgakis,
{\em {Nonabelian (2,0) tensor multiplets and 3-algebras},}
\href{http://dx.doi.org/10.1007/JHEP08(2010)083}{JHEP {\bf 08} (2010) 083}
  [{\tt \href{http://www.arxiv.org/abs/1007.2982}{1007.2982 [hep-th]}}].

\bibitem{Kawamoto:2011ab}
S.~Kawamoto, T.~Takimi, and D.~Tomino,
{\em {Branes from non-Abelian (2,0) tensor multiplet with 3-algebra},}
{\tt \href{http://www.arxiv.org/abs/1103.1223}{1103.1223 [hep-th]}}.

\bibitem{Honma:2011br}
Y.~Honma, M.~Ogawa, and S.~Shiba,
{\em {Dp-branes, NS5-branes and U-duality from nonabelian (2,0) theory with Lie
  3-algebra},}
{\tt \href{http://www.arxiv.org/abs/1103.1327}{1103.1327 [hep-th]}}.

\bibitem{Freund:1981qw}
P.~G.~Freund and R.~I.~Nepomechie,
{\em {Unified geometry of antisymmetric tensor gauge fields and gravity},}
\href{http://dx.doi.org/10.1016/0550-3213(82)90356-X}{Nucl. Phys. B {\bf 199}
  (1982) 482}.

\bibitem{0817647309}
J.-L.~Brylinski,
{\em Loop spaces, characteristic classes and geometric quantization,}
Birkh{\"a}user Boston (2007).

\bibitem{Saemann:2010cp}
C.~Saemann,
{\em {Constructing self-dual strings},}
accepted by Commun.\ Math.\ Phys.
{\tt \href{http://www.arxiv.org/abs/1007.3301}{1007.3301 [hep-th]}}.

\bibitem{Gustavsson:2005fp}
A.~Gustavsson,
{\em {A reparametrization invariant surface ordering},}
\href{http://dx.doi.org/10.1088/1126-6708/2005/11/035}{JHEP {\bf 11} (2005)
  035} [{\tt \href{http://www.arxiv.org/abs/hep-th/0508243}{hep-th/0508243}}].

\bibitem{Gustavsson:2005aq}
A.~Gustavsson,
{\em {The non-abelian tensor multiplet in loop space},}
\href{http://dx.doi.org/10.1088/1126-6708/2006/01/165}{JHEP {\bf 01} (2006)
  165} [{\tt \href{http://www.arxiv.org/abs/hep-th/0512341}{hep-th/0512341}}].

\bibitem{Gustavsson:2008dy}
A.~Gustavsson,
{\em {Selfdual strings and loop space Nahm equations},}
\href{http://dx.doi.org/10.1088/1126-6708/2008/04/083}{JHEP {\bf 04} (2008)
  083} [{\tt \href{http://www.arxiv.org/abs/0802.3456}{0802.3456 [hep-th]}}].

\bibitem{Kawamoto:2000zt}
S.~Kawamoto and N.~Sasakura,
{\em {Open membranes in a constant $C$-field background and noncommutative
  boundary strings},}
\href{http://dx.doi.org/10.1088/1126-6708/2000/07/014}{JHEP {\bf 07} (2000)
  014} [{\tt \href{http://www.arxiv.org/abs/hep-th/0005123}{hep-th/0005123}}].

\bibitem{Bergshoeff:2000jn}
E.~Bergshoeff, D.~S.~Berman, J.~P.~van~der~Schaar, and P.~Sundell,
{\em {A noncommutative M-theory five-brane},}
\href{http://dx.doi.org/10.1016/S0550-3213(00)00476-4}{Nucl. Phys. B {\bf 590}
  (2000) 173} [{\tt
  \href{http://www.arxiv.org/abs/hep-th/0005026}{hep-th/0005026}}].

\bibitem{Huang:2010db}
K.-W.~Huang and W.-H.~Huang,
{\em {Lie 3-algebra non-abelian (2,0) theory in loop space},}
{\tt \href{http://www.arxiv.org/abs/1008.3834}{1008.3834 [hep-th]}}.

\bibitem{Basu:2004ed}
A.~Basu and J.~A.~Harvey,
{\em The M2-M5 brane system and a generalized Nahm's equation,}
\href{http://dx.doi.org/10.1016/j.nuclphysb.2005.02.007}{Nucl. Phys. B {\bf
  713} (2005) 136} [{\tt
  \href{http://www.arxiv.org/abs/hep-th/0412310}{hep-th/0412310}}].

\bibitem{Gustavsson:2006ie}
A.~Gustavsson,
{\em {Loop space, (2,0) theory, and solitonic strings},}
\href{http://dx.doi.org/10.1088/1126-6708/2006/12/066}{JHEP {\bf 12} (2006)
  066} [{\tt \href{http://www.arxiv.org/abs/hep-th/0608141}{hep-th/0608141}}].

\bibitem{Nahm:1979yw}
W.~Nahm,
{\em {A simple formalism for the BPS monopole},}
\href{http://dx.doi.org/10.1016/0370-2693(80)90961-2}{Phys. Lett. B {\bf 90}
  (1980) 413}.

\bibitem{Nahm:1981nb}
W.~Nahm,
{\em {All selfdual multi-monopoles for arbitrary gauge groups},}
Presented at Int. Summer Inst. on Theoretical Physics, Freiburg, West Germany,
  Aug 31 - Sep 11, 1981.

\bibitem{Nahm:1982jb}
W.~Nahm,
{\em {The construction of all selfdual multi-monopoles by the ADHM method},}
talk at the Meeting on Monopoles in Quantum Field Theory, ICTP, Trieste (1981).

\bibitem{Filippov:1985aa}
V.~T.~Filippov,
{\em $n$-Lie algebras,}
\href{http://dx.doi.org/10.1007/BF00969110}{Sib. Mat. Zh. {\bf 26} (1985) 126}.

\bibitem{deAzcarraga:2010mr}
J.~A.~de~Azcarraga and J.~M.~Izquierdo,
{\em {n-ary algebras: a review with applications},}
\href{http://dx.doi.org/10.1088/1751-8113/43/29/293001}{J. Phys. A {\bf 43}
  (2010) 293001} [{\tt \href{http://www.arxiv.org/abs/1005.1028}{1005.1028
  [math-ph]}}].

\bibitem{Gomis:2008uv}
J.~Gomis, G.~Milanesi, and J.~G.~Russo,
{\em {Bagger-Lambert theory for general Lie algebras},}
\href{http://dx.doi.org/10.1088/1126-6708/2008/06/075}{JHEP {\bf 06} (2008)
  075} [{\tt \href{http://www.arxiv.org/abs/0805.1012}{0805.1012 [hep-th]}}].

\bibitem{Benvenuti:2008bt}
S.~Benvenuti, D.~Rodriguez-Gomez, E.~Tonni, and H.~Verlinde,
{\em {$\CN=8$ superconformal gauge theories and M2 branes},}
\href{http://dx.doi.org/10.1088/1126-6708/2009/01/078}{JHEP {\bf 01} (2009)
  078} [{\tt \href{http://www.arxiv.org/abs/0805.1087}{0805.1087 [hep-th]}}].

\bibitem{Ho:2008ei}
P.-M.~Ho, Y.~Imamura, and Y.~Matsuo,
{\em {M2 to D2 revisited},}
\href{http://dx.doi.org/10.1088/1126-6708/2008/07/003}{JHEP {\bf 07} (2008)
  003} [{\tt \href{http://www.arxiv.org/abs/0805.1202}{0805.1202 [hep-th]}}].

\bibitem{DeMedeiros:2008zm}
P.~De~Medeiros, J.~M.~Figueroa-O'Farrill, and E.~Mendez-Escobar,
{\em {Lorentzian Lie 3-algebras and their Bagger-Lambert moduli space},}
\href{http://dx.doi.org/10.1088/1126-6708/2008/07/111}{JHEP {\bf 07} (2008)
  111} [{\tt \href{http://www.arxiv.org/abs/0805.4363}{0805.4363 [hep-th]}}].

\bibitem{Douglas:2010iu}
M.~R.~Douglas,
{\em {On D=5 super Yang-Mills theory and (2,0) theory},}
{\tt \href{http://www.arxiv.org/abs/1012.2880}{1012.2880 [hep-th]}}.

\bibitem{Lambert:2010iw}
N.~Lambert, C.~Papageorgakis, and M.~Schmidt-Sommerfeld,
{\em {M5-branes, D4-branes and quantum 5D super-Yang-Mills},}
\href{http://dx.doi.org/10.1007/JHEP01(2011)083}{JHEP {\bf 1101} (2011) 083}
  [{\tt \href{http://www.arxiv.org/abs/1012.2882}{1012.2882 [hep-th]}}].

\bibitem{Howe:1997ue}
P.~S.~Howe, N.~D.~Lambert, and P.~C.~West,
{\em The self-dual string soliton,}
\href{http://dx.doi.org/10.1016/S0550-3213(97)00750-5}{Nucl. Phys. B {\bf 515}
  (1998) 203} [{\tt
  \href{http://www.arxiv.org/abs/hep-th/9709014}{hep-th/9709014}}].

\bibitem{Mukhi:2008ux}
S.~Mukhi and C.~Papageorgakis,
{\em {M2 to D2},}
\href{http://dx.doi.org/10.1088/1126-6708/2008/05/085}{JHEP {\bf 05} (2008)
  085} [{\tt \href{http://www.arxiv.org/abs/0803.3218}{0803.3218 [hep-th]}}].

\bibitem{Constable:1999ac}
N.~R.~Constable, R.~C.~Myers, and O.~Tafjord,
{\em The noncommutative BIon core,}
\href{http://dx.doi.org/10.1103/PhysRevD.61.106009}{Phys. Rev. D {\bf 61}
  (2000) 106009} [{\tt
  \href{http://www.arxiv.org/abs/hep-th/9911136}{hep-th/9911136}}].

\bibitem{Nastase:2009ny}
H.~Nastase, C.~Papageorgakis, and S.~Ramgoolam,
{\em {The fuzzy $S^2$ structure of M2-M5 systems in ABJM membrane theories},}
\href{http://dx.doi.org/10.1088/1126-6708/2009/05/123}{JHEP {\bf 05} (2009)
  123} [{\tt \href{http://www.arxiv.org/abs/0903.3966}{0903.3966 [hep-th]}}].

\bibitem{Cherkis:2008qr}
S.~Cherkis and C.~Saemann,
{\em {Multiple M2-branes and generalized 3-Lie algebras},}
\href{http://dx.doi.org/10.1103/PhysRevD.78.066019}{Phys. Rev. D {\bf 78}
  (2008) 066019} [{\tt \href{http://www.arxiv.org/abs/0807.0808}{0807.0808
  [hep-th]}}].

\bibitem{Palmer:2011vx}
S.~Palmer and C.~Saemann,
{\em {Constructing generalized self-dual strings},}
{\tt \href{http://www.arxiv.org/abs/1105.3904}{1105.3904 [hep-th]}}.

\end{thebibliography}

\end{document}